\documentclass[structabstract]{aa}
\usepackage{amsmath}
\usepackage{graphicx}
\usepackage{txfonts}

\makeatletter
    
    \newcommand{\Rmnum}[1]{\expandafter\@slowromancap\romannumeral #1@}
\makeatother

\begin{document}

\graphicspath{{Figures/}}
   \title{Calibration of the angular momenta of the minor planets in the solar system}

   \author{Jian Li$^{1}$,
         Zhihong Jeff Xia$^{2}$, Liyong Zhou$^{1}$           
          }

   \institute{$^{1}$School of Astronomy and Space Science \& Key Laboratory of Modern Astronomy and Astrophysics in Ministry of Education,\\
~Nanjing University, 163 Xianlin Road, Nanjing 210023, PR China\\
\email{ljian@nju.edu.cn}\\
$^{2}$Department of Mathematics, Northwestern University, 2033 Sheridan Road, Evanston, IL  60208, USA\\
    \email{xia@math.northwestern.edu}
             }

   \date{\today}

  \abstract
 {}
   {We aim to determine the relative angle between the total angular momentum of the minor planets and that of the Sun-planets system, and to improve the orientation of the invariable plane of the solar system.}
 {By utilizing physical parameters available in public domain archives, we assigned reasonable masses to 718041 minor planets throughout the solar system, including near-Earth objects, main belt asteroids, Jupiter trojans, trans-Neptunian objects, scattered-disk objects, and centaurs. Then we combined the orbital data to calibrate the angular momenta of these small bodies, and evaluated the specific contribution of the massive dwarf planets. The effects of uncertainties on the mass determination and the observational incompleteness were also estimated.}
   {We determine the total angular momentum of the known minor planets to be $1.7817\times10^{46}$ g $\cdot$ cm$^2$ $\cdot$ s$^{-1}$. The relative angle $\alpha$ between this vector and the total angular momentum of the Sun-planets system is calculated to be about $14.74^\circ$. By excluding the dwarf planets Eris, Pluto, and Haumea, which have peculiar angular momentum directions, the angle $\alpha$ drops sharply to $1.76^\circ$; a similar result applies to each individual minor planet group (e.g., trans-Neptunian objects). This suggests that, without these three most massive bodies, the plane perpendicular to the total angular momentum of the minor planets would be close to the invariable plane of the solar system. On the other hand, the inclusion of Eris, Haumea, and Makemake can produce a difference of 1254 mas in the inclination of the invariable plane, which is much larger than the difference of 9 mas induced by Ceres, Vesta, and Pallas as found previously. By taking into account the angular momentum contributions from all minor planets, including the unseen ones, the orientation improvement of the invariable plane is larger than 1000 mas in inclination with a $1\sigma$ error of $\sim50-140$ mas. }
   {}


   \maketitle
%

\section{Introduction}

The solar system was born in a giant rotating cloud of gas and dust, known as the solar nebular, which determines its initial angular momentum. After the solar nebular collapsed, the central part transformed into the proto-Sun, and the outer part gave rise to the eight planets and numerous minor planets. Through some physical and dynamical processes, the proto-Sun could have lost most of its angular momentum, but only a small fraction of the ejected part entered the planetary system. In the present day, the Sun, with 99.9\% of the mass of the entire solar system, only has less than 0.6\% of the total angular momentum, while planets with $<1\%$ of the total mass possess more than 99\% of the angular momentum (Dai 1977). Nevertheless, previous studies have not yet comprehensively assessed the angular momenta of the minor planets in the solar system.

The planets and the minor planets were both formed in a circumsolar nebular disk where the seeds had nearly coplanar orbits, thus the initial total angular momentum vectors of these two populations should have aligned directions that were perpendicular to the disk plane. However, there are a great many samples of minor planets discovered to be on high-inclination orbits throughout the solar system (Li et al. 2014a, b). According to the current observation, more than $\sim45$\% of the main belt asteroids (MBAs) are moving on orbits with inclinations $i>10^\circ$. This percentage goes up to $\sim52$\% for the population of trans-Neptunian objects (TNOs), scattered-disk objects (SDOs),\footnote{The Minor Planet Center separates the TNOs in stable orbits from the SDOs in scattered (unstable) orbits. Nevertheless, the distinction between these two populations may no be clear cut.} and centaurs. Furthermore, among these minor planets in the outer solar system, there are about one hundred extremely inclined ones with $i>40^\circ$, and some are even found on retrograde orbits ($i>90^\circ$), such as 2008 KV$_{42}$ with $i=103^\circ$ (Gladman et al. 2009) and 2011 KT$_{19}$ with $i=110^\circ$ (Chen et al. 2016). We are aware that the high-inclination orbit would significantly tilt the direction of an object's angular momentum vector. However, if we take into account all the minor planets with a very wide inclination dispersion, the direction of their total angular momentum remains uncertain. This direction could be crucial since it may affect the invariable plane of the solar system, which is defined as the plane perpendicular to the system's total angular momentum vector.

Nowadays, the notable high-inclination minor planets put forward a great challenge for the dynamical evolution of the solar system. Various scenarios have been proposed to account for the inclination excitation, for example, stellar encounter (Ida et al. 2000; Kobayashi et al. 2005), large scattered planetesimals (Petit et al. 1999; Lykawka \& Mukai 2008), and Planet 9 (Batygin \& Brown 2016). In the view of angular momentum transportation, these three mechanisms work in inhomogeneous backgrounds, so they may cause the direction of the total angular momentum of the minor planets away from that of the planets. Nevertheless, the minor planets would exchange angular momentum with the planets through secular interactions (Murray \& Dermott 1999), and their angular momentum vectors process on the order of, or shorter than, 10 Myr in the solar system. Then it may be expected that any external perturbation that might impart an asymmetrical change in the overall angular momentum of the minor planets would be distributed to the entire planetary system over a relatively short timescale. As proposed by Volk \& Malhotra (2017), this relaxation timescale should be comparable to the mentioned $\lesssim10$ Myr procession timescale of the angular momentum vectors. If this is the case, then the relative angle should not be significant between the total angular momentum vector of the minor planets and that of the planets. One must bear in mind that the classical disturbing function for secular perturbations in the three-body problem is expanded with respect to small inclination, which is not adequate for high inclination. To support the above theoretical argument, a complete and reasonable measurement of the total angular momentum of the minor planets is warranted.

A few astronomers have worked on determining the mid-plane (i.e., the average orbital plane) of the minor planets. Collander-Brown et al. (2003) focused on the classical TNOs with semimajor axes between 40 and 47 AU, while the samples were only 141, and the biggest two components (Haumea and Makemake) were not discovered at that time. They did mention that their analysis may be incorrect if there is an unseen Pluto-size object. As for the MBAs, Cambioni \& Malhotra (2018) recently considered a nearly complete and unbiased sample. They computed the average of unit angular momentum vectors of this subset to measure the mid-plane of the main belt and compare with the prediction of secular perturbation theory. In this paper, we aim to calibrate the angular momenta of all the minor planets in the present solar system, including near-Earth objects (NEOs), MBAs, Jupiter trojans (JTs), TNOs, SDOs, and centaurs. In this way, we can calculate the plane perpendicular to the total angular momentum of the minor planets, which is different from the mid-plane studied in previous works. Therefore, the masses of the minor planets must be included in the set of parameters, since the massive objects may dominate the total angular momentum of the entire population, not only the modulus but also the direction. Our samples have well-determined orbital elements, and some available physical parameters (e.g., diameters and albedos) in public domain archives. By analyzing the observational data and using some assumptions, we intend to make a reasonable evaluation of the minor planets' masses, and consequently carry out a reasonably accurate measurement of their total angular momentum. We would also like to make a comparison between the different groups of minor planets.

Additionally, we may be able to improve the orientation of the invariable plane of the solar system. Comparing to the basic system including the Sun, the eight planets and the dwarf planet Pluto, Souami \& Souchay (2012) evaluated the effects of the dwarf planet Ceres as well as the other two biggest MBAs (Vesta and Pallas) on the determination of the invariable plane. They showed that, for the DE405 ephemeris, the inclusion of these three bodies can change the inclination of the invariable plane with respect to the equinox-ecliptic by 9 milliarcseconds (mas). Since it is so, one can expect that the remaining three dwarf planets (Eris, Haumea, and Makemake) should also be taken into account due to their much larger angular momenta, that is, they can induce non-negligible variation in the invariable plane's orientation. Furthermore, it is worth estimating the possible contribution from the rest of the numerous minor planets.

The remainder of this paper is organized as follows. In Sect. 2, we build a database containing orbital elements and reasonable masses of 718041 minor planets throughout the solar system. In Sect. 3, we calculate the total angular momentum vectors of the Sun-planets system and the cataloged minor planets, respectively, and analyze the angle between these two vectors. Also, we make the uncertainty estimation of the derived relative angle. Using the obtained angular momenta for individual minor planets, in Sect. 4, we further investigate the orientation of the invariable plane of the solar system and present the associated improvement. Our conclusions and discussion are given in Sect. 5.


\section{Minor planet database}

We selected the minor planets from the Asteroid Orbital Elements Database (ASTORB)\footnote{ftp://ftp.lowell.edu/pub/elgb/astorb.html} as our samples in the database, including NEOs, MBAs, JTs, TNOs, SDOs, and centaurs. As of 2016 October, the total number of these populations with high-precision orbital elements amounts to 718041. The ASTORB file also supplies some physical parameters such as the absolute magnitude; however, only a very small fraction of the objects have measured diameters, albedos, and IRAS (Infrared Astronomical Satellite) taxonomic classifications, which are essential to determine the masses. Only if an object's mass is reasonably assigned will it be possible to calculate the real, but not the unit, angular momentum.

\subsection{Mass determination}

\subsubsection{Objects with measured masses}

\begin{table*}
\centering
\begin{minipage}{13cm}
\caption{Eleven largest minor planets with diameters $D>800$ km.}      
\label{bigs}
\begin{tabular}{l l l l l l}        
\hline                 
Number    &      Name       & Mass ($10^{-10}M_{\odot}$) &    Group                                  &      Reference                  \\

\hline
 
1                &   Ceres                &         $4.756\pm0.004$                                    &   MBA; dwarf planet         &    Fienga et al. (2008)           \\

50000       &   Quaoar   &         $7.038\pm1.005$                      &   TNO                                    &    Fraser et al. (2013)        \\

90377       &   Sedna\footnote{Density is assumed to be the same as Eris, $2.3\pm0.3$ g/cm$^3$, since both are SDOs and dwarf planets.}              &         5.964$^{+2.538}_{-1.931}$      &   SDO                     &    P\'al et al. (2012)                             \\

90482       &   Orcus              &         $3.223\pm0.096$                 &   TNO                                      &    Carry et al. (2011)    \\

120347     &   Salacia             &         $2.343\pm0.111$                  &   TNO                                     &    Stansberry et al. (2012)  \\

134340     &   Pluto               &       $73.504\pm0.211$                  &   TNO; dwarf planet         &    Stern et al. (2015)               \\

136108     &   Haumea    &      $20.239\pm0.214$                   &   TNO; dwarf planet        &    Ragozzine \& Brown (2009)   \\

136199     &   Eris                   &       $83.455\pm1.005$                   &   SDO; dwarf planet     &    Brown \& Schaller (2007)      \\

136472     &   Makemake  &       13.086$^{+2.602}_{-2.512}$                                     &   TNO; dwarf planet        &    Ortiz et al. (2012)    \\

225088     &   2007 OR10         &       14.282$^{+2.197}_{-5.405}$                                     &   SDO                                   &    P\'al et al. (2016)                     \\

307261     &   2002 MS4\footnote{Density is assumed to be the same as Salacia, $1.29^{+0.29}_{-0.23}$ g/cm$^3$ (Fornasier et al. 2013), since both are TNOs trapped in Neptune's 1:2 mean motion resonance.}         &         2.767$^{+1.160}_{-0.820}$       &   TNO                                    &   Vilenius et al. (2012)            \\
\hline
\end{tabular}
\end{minipage}
\end{table*}

In the solar system, the 11 largest minor planets have diameters of $D>800$ km, including the 5 dwarf planets (Ceres, Pluto, Haumea, Makemake, and Eris). The most updated masses of these objects are given in Table \ref{bigs}. 

Considering smaller minor planets that have diameters of 40 km $<D<$ 800 km, to the best of our knowledge, we managed to collect masses for 268 objects from the published literature. Among them, there are only 33 ($\sim$12.3\%) populated in the outer solar system, while the TNOs of some particular sizes are likely to be much more than the MBAs. For instance, the main belt holds around 200 asteroids larger than 100 km in diameter, which have absolute magnitudes $H<10$ and are supposed to be observationally complete, while the number of such large objects in the trans-Neptunian region between 30 and 50 AU could be as many as $\sim10^5$ (Trujillo et al. 2001; Brunini 2002). Thus for our samples with known masses in the outer solar system, the low proportion is due to the fact that the larger distance generally makes an object fainter and more difficult to be directly measured. Actually most of them are binaries, and we use their system masses that could be calculated from the orbital periods and semimajor axes by applying Kepler's third law.

Taken in total, we now have 279 minor planets with measured masses, and they were firstly added to our database.

\subsubsection{Objects with measured diameters}

To date, the Wide-field Infrared Survey Explorer (WISE) observations have provided the most complete and accurate measurements of diameters and albedos ($P_v$) for the minor planet population. This project observed mostly targets interior to or near the orbit of Jupiter, of which there were over 130000 MBAs (Masiero et al. 2011; Grav et al. 2012a; Masiero et al. 2012), about 700 NEOs (Mainzer et al. 2011, 2012, 2014), and approximately 1900 JTs (Grav et al. 2011, 2012b). However, for the more distant TNOs, SDOs, and centaurs (here after TSCs for short), they are much more difficult to detect at infrared wavelengths. In the WISE survey only 52 SDOs and centaurs have been observed, and the mean of their heliocentric distances at the time of observation is as small as 10.5 AU (Bauer et al. 2013).

Unfortunately, a vast majority of the TSCs are too cold (down to $\sim35$ K) to be spotted by the WISE. Among these icy bodies, we chose the objects with available $D$ and $P_v$ values that are either from Mike Brown's website\footnote{http://web.gps.caltech.edu/$\sim$mbrown/dps.html} or documented in the series of eleven `TNOs are Cool' papers (Duard et al. 2010; Lellouch et al. 2010; Lim et al. 2010; M\"uller et al. 2010; Mommert et al. 2012; Santos-Sanz et al. 2012; Vilenius et al. 2012; Fornasier et al. 2013; Lellouch et al. 2013; P\'al et al. 2012; Vilenius et al. 2014).

For all the minor planets with measured diameters considered here, to assign them with reasonable masses, we evaluated their densities into four different groups:\\

\noindent\textit{--MBAs}

The (optical) albedo is a key parameter for the surface property of a celestial body, therefore it is helpful to access the density. We borrowed the idea from the statistical model proposed by Tedesco et al. (2005), who parameterized the MBAs with $D>1$ km into four albedo classes as shown in Table \ref{class}. In the WISE data, diameters of the MBAs are computed from the infrared thermal flux, then albedos are determined by combining diameters with  literature values of absolute magnitudes $H$ (Masiero et al. 2011). However, a small fraction ($\sim1$\%) of the MBAs with measured diameters have no available $H$ values due to various reasons (Nugent et al. 2015). So we incorporate $H$ from the ASTORB file to estimate their albedos $P_v$ with the formula
\begin{equation}
P_v=\left(\frac{1329\times10^{-H/5}}{D}\right)^2.
\label{albedo}
\end{equation}
For each MBA with a given $P_v$, we can accordingly assign it to one of the three density classes in Table \ref{class}: C-type (low albedo), M-type (intermediate or high albedo), and S-type (moderate albedo) (Britt et al. 2002). The mean densities and corresponding uncertainties for the C-, M-, S-asteroids are taken from the INPOP08 (French Int\'egrateur Num\'erique Plan\'etaire de l'Observatoire de Paris) ephemeris (Fienga et al. 2009). Although asteroid masses have been updated since the INPOP08, in particular those obtained with the INPOP17a (Viswanathan et al. 2017), there would be probably a minor effect.

\begin{table}
\centering
\hspace{-45pt}
\begin{minipage}{8cm}
\caption{Definitions of albedo ($P_v$) and density ($\rho$) classes for the MBAs.}      
\label{class}
\begin{tabular}{ l c c c}       
\hline                 
Albedo class    &        $P_v$            &         Density class      &   $\rho$ (g/cm$^3$)  \\

\hline
 
Low               &     0.020 -- 0.089      &         C-type                 &     1.54 $\pm$ 0.07      \\

Intermediate    &     0.089 -- 0.112       &         M-type                   &     4.98 $\pm$ 0.50        \\

Moderate         &      0.112 -- 0.355   &         S-type                     &     1.94 $\pm$ 0.14         \\

High                    &      0.355 -- 0.526    &         M-type                     &     4.98 $\pm$ 0.50           \\

\hline
\end{tabular}
\end{minipage}
\end{table}

For our nominal case, the class densities are fixed to be the mean values: $\rho=1.54$ g/cm$^3$ [C-type], 4.98 g/cm$^3$ [M-type], and 1.94 g/cm$^3$ [S-type]. Although the uncertainty of $\rho$ contributes to the improvement of planetary ephemerides (Kuchynka et al. 2010), it does not affect the conclusions of this paper. A series of additional tests, with marginally distinct $\rho$ from the mean values in the density intervals given in Table \ref{class}, have been carried out. We find that the outcomes are exactly the same as those of the nominal case.\\

\noindent\textit{--NEOs}

Minor planets are named NEOs if their closest approaches to the Sun (i.e., perihelia) are less than 1.3 AU. Since the NEOs are believed to originate from the fragments of the MBAs (Fern\'andez et al. 2014; Granvik et al. 2017), these two groups should share similar physical features. By repeating the same procedure used above, we can also assign each NEO to a class density according to its available albedo.\\

\noindent\textit{--JTs} 

Here, we adopt a default and typical density of $\rho=2$ g/cm$^3$ for this trojan population, as proposed by Jewitt et al. (2000).\\

\noindent\textit{--TSCs} 

The SDOs are minor planets with eccentricities as high as 0.8, and they are also known as scattered TNOs. The centaurs are those objects with semimajor axes and perihelia between the orbits of Jupiter and Neptune, and they are generally thought to be escaped TNOs. Both of them probably initially inhabited the region where the primordial TNOs were born, and then they were delivered to their current locations (Levison et al. 2008). In addition, just like the TNOs, the SDOs and the centaurs also appear to be ice-dominated objects according to spectral analysis. Thus we may assume that these three populations have the similar physical properties (e.g., density), and could be considered as a combined group: the TSCs.

Up to now, there are 40 TSCs among the 279 samples, with measured masses collected in Sect. 2.2.1. They have estimated densities ranging from $\rho=0.5$ to 1.5 g/cm$^3$ (Vilenius et al. 2012, 2014; Carry 2012). These objects are relatively large and have diameters of approximately $D=100-1000$ km, while for the smaller ones with diameters down to $D\sim1$ km, their densities remain quite uncertain. A recent study indicates that the common low-density ice is methane, with $\rho=0.5$ g/cm$^3$, implying a lower density limit for the TSCs (Vilenius et al. 2014). Based on these works, we decided to adopt the mid-value of $\rho=1$ g/cm$^3$, which actually is the density of liquid water and may apply to the small TSCs composed of icy pieces (Parker et al. 2011).\\

Usage of the observed diameter $D$ and the assigned density $\rho$ has allowed us to calculate the mass $M$ of an object through the equation
\begin{equation}
M=\frac{4}{3}\pi\cdot\left(\frac{D}{2}\right)^3\cdot\rho~.
\label{mass}
\end{equation}
In this way, we have generated the masses for 135046 more minor planets, which are appended to our database.

\subsubsection{Objects with only absolute magnitudes}

Among the minor planets listed in the ASTORB catalog, masses have not yet been determined for more than 80\% of them. These objects have neither available diameters nor albedos from the WISE survey, because they are too faint to be observed with infrared spectroscopy. Then the only physical parameter we can accurately access is the absolute magnitude. Indeed, the minor planets from different groups may have distinct albedos associated to the formation and evolution processes in the early solar system. Given a plausible value of the albedo $P_v$, the diameter $D$ can be converted from the absolute magnitude $H$ by the deformation formula of Eq. (\ref{albedo}),
\begin{equation}
D=\frac{1329}{\sqrt{P_v}}\cdot10^{-H/5}.
\label{diameter}
\end{equation}
Then, the density and subsequent mass determinations can proceed just as we have done for the samples with measured diameters.

We have argued that the MBAs and the NEOs probably share similar surface properties. Accordingly, a total number of 577150 more objects from these two groups are randomly assigned one of the four albedo classes with probabilities deduced by Kuchynka et al. (2010): 0.56 for low albedo, 0.07 for intermediate albedo, 0.34 for moderate albedo, and 0.03 for high albedo.  Using the assumed albedo $P_v$, the density $\rho$ can be given corresponding to a specific albedo class (see Table \ref{class}).

As in the Jupiter trojan model that we recently built (Li \& Sun 2018), we chose a fixed albedo of $P_v=0.04$ for the remaining 4744 JTs. Observations of small JTs revealed that this $P_v$ is typical according to their albedo distribution (Jewitt et al. 2000; Fern\'andez et al. 2003; Nakamura \& Yoshida 2008; Yoshida \& Nakamura 2008). The density is still assumed to be the same, $\rho=2$ g/cm$^3,$ as we used in Sect. 2.1.2.

Regarding the TSCs, the Spitzer Space Telescope observations derived a mean albedo of $P_v=0.07-0.08$ (Stansberry et al. 2008), which is almost identical with the later estimation from the program `TNOs are Cool' based on a much larger sample size (Mommert et al. 2012). Within this $P_v$ interval, we chose albedos randomly for the remaining 822 TSCs. As we discussed before, their densities are still adopted to be that of liquid water, $\rho=1$ g/cm$^3$.

So far, in this subsection we have assigned plausible masses to an additional 582716 minor planets. Since absolute magnitudes of these objects may be often not very accurate, we assume an offset of $\sim0.2-0.4$ mag as reported by some authors (Juri\'c et al. 2002; Fraser et al. 2014).  With Eq. (\ref{diameter}), this introduces a 10\%-20\% uncertainty directly on the diameter, resulting in only a $<1$\% uncertainty on the minor planet's mass. As for the mass uncertainty introduced by the assumed bulk densities, it is about less than 10\% for the MBAs, but could possibly be larger for the JTs and the TSCs due to scarce data on their surface properties. Nevertheless, all the minor planets considered here are rather small, and so they have little contribution to the total angular momentum of the overall population, as we will show in Sect. 3.2.

\subsection{Data combination}

We combine the three sets of minor planets with assigned masses in Sect. 2.1, and then crossmatch them against the orbital data from the ASTORB file. Finally, we have constructed our own database containing the masses and orbital elements of 718041 minor planets throughout the solar system, including four individual groups: 14912 NEOs, 693949 MBAs, 6587 JTs, and 2593 TSCs. These objects would be registered as nominal samples in this work.    


\section{Angular momentum in the solar system}

\subsection{Overall results}

For the fundamental Sun-planets system, the total angular momentum $\vec{H}_{S\&p}$ relative to the solar system barycenter can be calculated by
\begin{equation}
\vec{H}_{S\&p}=m_{\odot}\cdot\vec{r}_{\odot}\times\vec{v}_{\odot}+\sum_{j=1}^{8}m_j\cdot\vec{r}_i\times\vec{v}_j,
\label{Hplanets}
\end{equation}
where $m$, $\vec{r,}$ and $\vec{v}$ are the mass, the barycentric position vector, and the barycentric velocity vector, respectively. The subscript $j$ refers to the planets from Mercury ($j=1$) to Neptune ($j=8$). The Sun and planets' masses, initial positions, and velocities are adopted from DE405 with epoch 1969 June 28 (Standish 1998). To be consistent with the ephemeris of minor planets in the ASTORB file, we transformed the positions and velocities of the Sun and planets from the mean equatorial system to the J2000.0 ecliptic system, and then numerically integrated their orbits to the epoch 2016 November 8. At this point, the absolute value of the total angular momentum of the Sun-planets system is computed to be $|\vec{H}_{S\&p}|=3.1333\times 10^{50}$ g $\cdot$ cm$^2$ $\cdot$ s$^{-1}$, which agrees well with the previously published results (Dai 1977; Wesson 1984; Weissman 1991).

Next, for each of the four minor planet groups, we separately calculated the overall angular momentum analogous to Eq. (\ref{Hplanets}). By summing by these vectors via
\begin{equation}
\vec{H}_{minor}=\vec{H}_{NEOs}+\vec{H}_{MBAs}+\vec{H}_{JTs}+\vec{H}_{TSCs},
\label{Hasteroids}
\end{equation}
we then carry out, for the first time, a reasonable measurement of the total angular momentum of the known minor planets in the solar system. The corresponding module is $|\vec{H}_{minor}|=1.7817\times10^{46}$ g $\cdot$ cm$^2$ $\cdot$ s$^{-1}$. It is noteworthy that the TSCs contribute about 98.7\% to the total angular momentum of the minor planets, and this fraction could go even higher due to the observational incompleteness of the TSC samples. The resultant measurement uncertainty in $\vec{H}_{minor}$ will be considered in later analysis.

\begin{figure}
 \hspace{0cm}
  \centering
  \includegraphics[width=9cm]{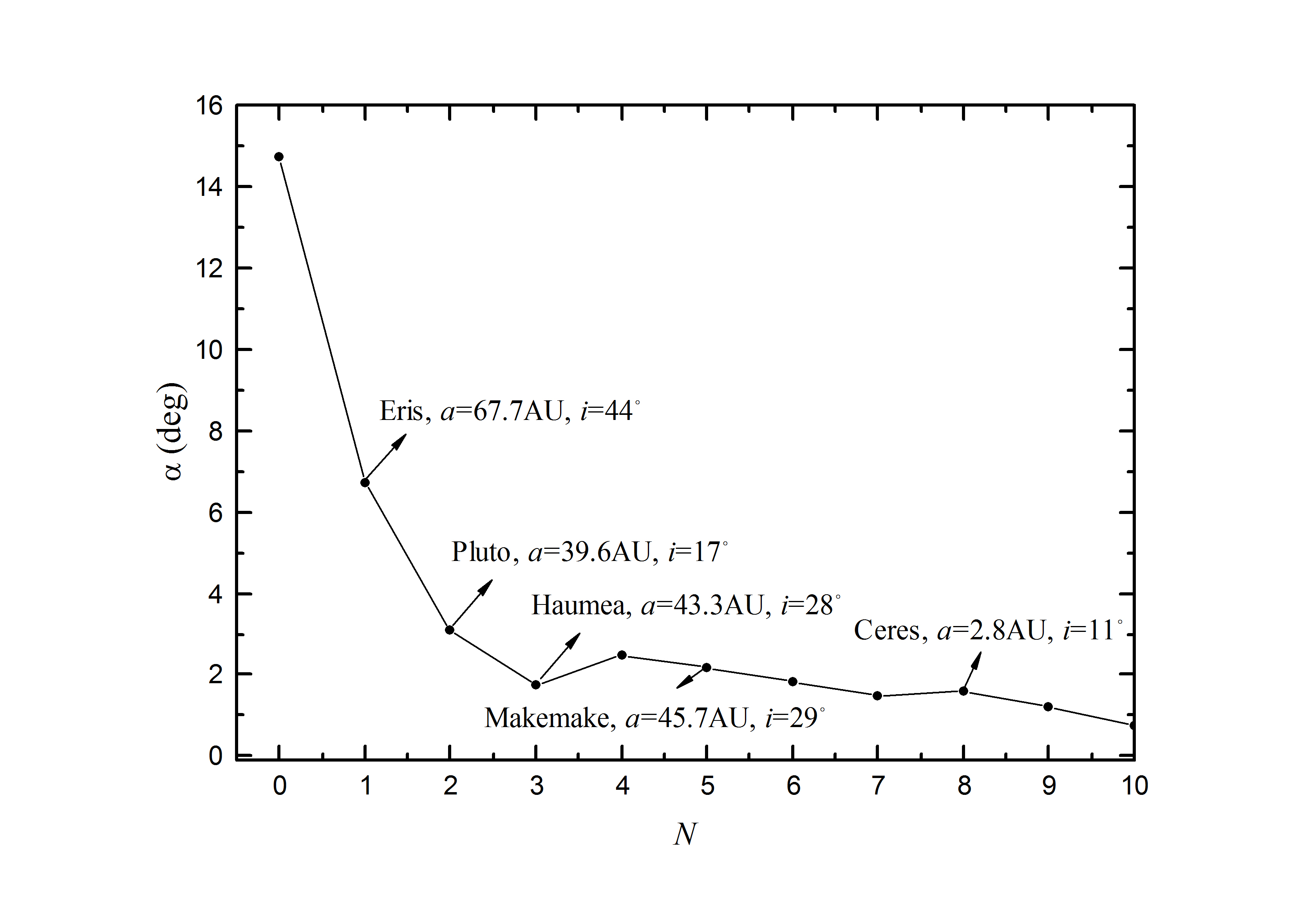}
  \caption{Variation of the relative angle $\alpha$ between the total angular momentum vectors of the Sun-planets system and the minor planets. The curve is plotted as the $N$ most massive minor planets have been excluded in our calculations, and each dot refers to a specific object with its semimajor axis ($a$) and inclination ($i$). The case of $N=0$ indicates that all the known minor planets have been taken into account. After ruling out the contamination of the three dwarf planets Eris, Pluto, and Haumea (i.e., $N=3$), $\alpha$ drops sharply and then maintains a very small value as $N$ continues to increase.}
  \label{alpha_vari}
\end{figure}

Using the obtained $\vec{H}_{S\&p}$ and $\vec{H}_{minor}$, we can derive the relative angle $\alpha$ between the total angular momentum vectors of the Sun-planets system and the minor planets from the equation
\begin{equation}
\alpha=\arccos\left(\frac{\vec{H}_{S\&p}\cdot\vec{H}_{minor}}{|\vec{H}_{S\&p}|\cdot|\vec{H}_{minor}|}\right).
\label{angle}
\end{equation}
This approach yields a quite large value of $\alpha\approx14.74^{\circ}$, which seems to contradict our theoretical expectation. As we said in the Introduction, due to the angular momentum exchange between the planets and the minor planets through secular interactions, even the two vectors $\vec{H}_{S\&p}$ and $\vec{H}_{minor}$ are not perfectly aligned with each other, and their relative angle $\alpha$ should not be so large. In fact, we have made some numerical experiments in the framework of the Sun, the eight plants, and hundreds of equal mass particles to mimic the TNOs. With different initial conditions for the particles, the systems are designed to start from $\alpha=20^{\circ}$. We find that the values of $\alpha$ do drop considerably after 10 Myr integrations, and then always oscillate below $\sim4^{\circ}$ during the subsequent several tens of Myr. We may thereby suppose that the angle $\alpha$ should be smaller or comparable to $4^{\circ}$. Alternatively, Collander-Brown et al. (2003) and Cambioni \& Malhotra (2018) have previously derived the mid-planes of the TNOs and the MBAs, respectively, which are perpendicular to the sum of the unit angular momentum vectors $\vec{r}\times\vec{v}$. Their results show that both mid-planes deviate from the invariable plane by less than $3^{\circ}$. Similarly, by assuming all our nominal samples have equal masses, we find that the corresponding relative angle, $\alpha$, will be only as small as $1.95^{\circ}$.

We then realized that such a large $\alpha\approx14.74^{\circ}$ is due to the contamination of the several biggest objects occupying high-inclination orbits, especially the dwarf planets Eris ($i=44.2^{\circ}$), Pluto ($i=17.1^{\circ}$), and Haumea ($i=28.2^{\circ}$). These objects not only carry a considerable portion of the angular momentum of the minor planets, but also have their angular momentum vectors deviated substantially from those of the planets. As a result, the authentic value of the angle $\alpha$ could be significantly misleading. Figure \ref{alpha_vari} presents the variation of $\alpha$ after removing the $N$ most massive minor planets. As we expected, by excluding Eris, Pluto, and Haumea (i.e., $N=3$), the angle $\alpha$ decreases dramatically to only about $1.76^{\circ}$ (i.e., $\alpha_2$ in the top row of Table \ref{alpha12}). When this procedure continues, the angle $\alpha$ will fluctuate slightly associated with the number $N$, and it can always maintain a small value. Since we have $\alpha\lesssim2^{\circ}$ for $N\geqslant3$, in these cases the directions of the total angular momenta of the Sun-planets system and the minor planets could be regarded as nearly coincident.

\begin{table}
\hspace{-0.5cm}
\centering
\begin{minipage}{9cm}
\caption{Relative angle between the total angular momentum vectors of the Sun-planets system and the minor planets; $\alpha_1$ and $\alpha_2$ are for the cases with and without the three dwarf planets Eris, Pluto, and Haumea, respectively. The last two columns refer to the changed values of $\alpha_2$ by taking into account some uncertainties, as we will illustrate in Sect. 3.2.}      
\label{alpha12}
\begin{tabular}{ c c c c c}       
\hline                 
Population                                  &    $\alpha_1(^{\circ})$    &     $\alpha_2(^{\circ})$    &     $\alpha_2'(^{\circ})$    &     $\alpha_2''(^{\circ})$   \\

\hline
 
NEOs+MBAs+JTs+TSCs              &               14.7449                 &                1.7635                     &                 1.8020                      &                  --                              \\

TSCs                                                &               14.9210                 &                1.8218                     &                 1.8612                      &                  --                               \\

TNOs                                               &                 9.7645                 &                1.9105                     &                 1.9651                      &                  --                                \\

SDOs                                               &               29.1900                  &                4.6759                     &                 4.6861                      &                  --                                    \\

MBAs                                               &                3.0922                  &                3.0922                     &                 3.1165                      &               3.0947                            \\

JTs                                                    &                3.1107           &                3.1107                     &                 3.4876                      &               3.1541                            \\            

\hline
\end{tabular}
\end{minipage}
\end{table}

Before the final conclusion is achieved, the investigation of focusing on the TSCs may allow us to be more confident about the above results. These icy objects have been wandering at the edge of the solar system, especially the SDOs, with aphelion distances as large as thousands of AU. If any asymmetric angular momentum transfer with interstellar space happened, the direction of their total angular momentum should suffer the greatest tilt. Table \ref{alpha12} lists the values of the angle $\alpha$, calculated with ($\alpha_1$) and without ($\alpha_2$) the three dwarf planets Eris, Pluto, and Haumea, for the TSCs and their subgroups. We find that the outcome for the TSC population is almost identical to that for the entire minor planet population (NEOs+MBAs+JTs+TSCs). As these three dwarf planets have been removed, the total angular momentum vectors of the Sun-planets system and the TSCs become nearly aligned to each other, yielding an angle of $\alpha_2\approx1.82^{\circ}$. Such a sharp drop of $\alpha$ also occurs when we consider the samples either from the subgroup TNOs or SDOs. Although the pole of the total angular momentum of the SDOs seems a bit more inclined by $\alpha_2\approx4.68^{\circ}$, this value is most likely due to the strong observational incompleteness of these extremely distant objects in the current survey. 

Table \ref{alpha12} also shows that the MBAs and the JTs both have small values of $\alpha_2\approx3.1^{\circ}$. For these two groups, the angles $\alpha_1$ and $\alpha_2$ are equal to each other because Eris, Pluto, and Haumea with $a>30$ AU are not their members.

In summary, the above study shows that, as long as the three dwarf planets Eris, Pluto, and Haumea are excluded, the relative angle ($\alpha_2$) between the directions of the total angular momenta of the Sun-planets system and the minor planets could be $\lesssim 4^{\circ}$. This rather small value seems consistent with the theoretical expectation based on secular perturbations, and our numerical experiments. Furthermore, the same also applies to individual minor planet groups. Henceforth we will refer to $\alpha_2$ as the authentic relative angle, and the measurement uncertainty of this angle will be discussed in the following.

\subsection{Uncertainty of the angle $\alpha_2$}

The first uncertainty of the deduced value of $\alpha_2$ is from the minor planets with only absolute magnitudes (see Sect. 2.1.3), which comprise more than 80\% of our nominal samples. We used a simple methodology to assign albedos for these objects, and then deduced their possible diameters. As a result, we may introduce quite a large error into the mass determination. In order to evaluate the resultant impact on the value of $\alpha_2$, we elected to remove all these faint objects from our nominal samples. The corresponding results are still restricted to the case of excluding Eris, Pluto, and Haumea, but denoted by the symbol $\alpha_2'$. 

As shown in the top row of Table \ref{alpha12}, for the entire minor planet population, the measured difference between the angles $\alpha_2$ and $\alpha_2'$ is only by about 2\%. It is easy to know the reason for this consistency:  among nominal samples, the objects with only absolute magnitudes are much smaller and have less than 0.7\% of the total mass, leading to as small as $\sim1.5$\% of the total angular momentum. Furthermore, for any individual group (e.g., TSCs) as shown in Table \ref{alpha12}, the difference between $\alpha_2$ and $\alpha_2'$ is also considerably small, on the order of a few percent. Therefore, in spite of the errors in the measurement of the angular momentum as concerned here, we believe that our main results would not be affected at all.

The second uncertainty could come from the adopted masses and orbits of the distant TSCs. To obtain a rough estimation, we first focus on the 41 TSCs with measured masses from Sect. 3.1.1. They have a total angular momentum of $|\vec{H}_{TSCs}^{41}|=1.2801\times10^{46}$ g $\cdot$ cm$^2$ $\cdot$ s$^{-1}$, which contributes a fraction of $\sim$73\% to $|\vec{H}_{TSCs}|$ for all the 2593 cataloged TSCs in our database. We note that, for some TSC samples, the mass determinations could be poor due to their distant orbits, as depicted by the largest minor planets in Table \ref{bigs}. Taking into account the mass uncertainty, we generated a set of ten random masses for each object within the error space. We then repeated the calculation of the total angular momentum of these 41 TSCs over the ten different mass sets, and the resulting variability in $|\vec{H}_{TSCs}^{41}|$ is $\lesssim2\%$. Besides the mass uncertainty, the orbits of the distant TSCs may often be not accurately determined. Therefore, to make the above assessment of $|\vec{H}_{TSCs}^{41}|$ more reliable, it is necessary to consider the bias of orbital parameters. Using the orbital elements and the associated 1-$\sigma$ variations provided by the AstDyS\footnote{http://hamilton.dm.unipi.it/astdys/}, we produced a population of 100 clones for each TSC in the six-dimensional orbit distribution. Given the standard masses, the changes in $|\vec{H}_{TSCs}^{41}|$ are found to be as small as $<0.03\%$. Taken in total, we suppose that the possible mass and orbit uncertainties are not likely to introduce a significant variation in the deduced angle $\alpha_2$ for the known TSCs.

The third uncertainty is the observational incompleteness. Cambioni \& Malhotra (2018) measured the mid-plane of the MBAs by the unit mean angular momentum, using the asteroid samples that are nearly complete at absolute magnitude up to $H=15.5$. According to this $H$ limit, all the fainter MBAs are removed from our nominal samples. Then we find that the angle $\alpha_2$ hardly changes, indicated by the resultant $\alpha_2''$ with a difference of only $\sim0.01^{\circ}$ (see Table \ref{alpha12}). This is because the largest members dominate the total angular momentum of the MBAs. For instance, the contribution of Ceres alone to the angle $\alpha_2$ measurement is about $0.64^{\circ}$ over the total of $\sim3.09^{\circ}$. For this bright subset of the MBAs with $H<15.5$, we further deduce that the plane perpendicular to their total angular momentum has an inclination of $i_J=4.7^{\circ}$ in the J2000.0 ecliptic system, which deviates greatly from the inclination $i_J=0.9^{\circ}$ obtained by Cambioni \& Malhotra (2018). This difference is plausibly due to our consideration of the asteroid mass in the angular momentum calculation, while Cambioni \& Malhotra (2018) used the unit vector. As a matter of fact, if we assume that all the samples are equal mass particles, the inclination $i_J$ would become a comparable value of $1.0^{\circ}$. 

A similar analysis for the JTs has also been carried out, by choosing the nearly complete samples with $H\le14$ (Li \& Sun 2018). As shown at the bottom of Table \ref{alpha12}, the resultant relative angle $\alpha_2''$ is about $3.15^{\circ}$, which is still very close to $\alpha_2=3.11^{\circ}$ for the entire JT population.

Unfortunately, the observational census of the TSCs is incomplete. This could be important because the TSCs dominate the total angular momentum of the minor planets, over 98\%, as we noticed in Sect. 3.1. The total mass of the known TSCs was estimated to be $\sim0.01M_{\oplus}$, while the intrinsic total mass (denoted by $M_{TSCs}$ hereafter) could be approximately $0.02M_{\oplus}$ (Fuentes \& Holman 2008; Fraser et al. 2014; Pitjeva \&  Pitjev 2018). Since there is no straightforward method to correct this incompleteness, Brown \& Pan (2004) and Volk \& Malhotra (2017) used indirect approaches to compute the mid-plane as the plane of symmetry of the TNOs' sky-plane motion vectors. They found that the mid-plane of the TNOs is close to the invariable plane of the solar system. Suppose that the spatial mass distribution of the complete TNO samples is nearly uniform, the plane perpendicular to the total angular momentum would not largely deviate from the mid-plane, thus it is also close to the solar system's invariable plane, meaning the angle $\alpha_2$ should be small. In order to explore how the observational incompleteness and biases in the cataloged TSCs affect our results, we  determine the possible angle $\alpha_2$ and the associated uncertainty by constructing Monte Carlo simulations.

For the ten known TSCs with $D>800$ km (see Table \ref{bigs}), this sample could be complete since no objects in such a size range have been discovered since mid-2007. These largest objects have a total mass of $\sim0.0075M_{\oplus}$, and they are held fixed in the Monte Carlo population when simulating the angular momentum measurements.

For the smaller TSCs with $D\le800$ km, which are far from observational completeness, we select the power-law size distribution as
\begin{equation}
N(D)\propto D^{-q},
\label{sizeDistribution}
\end{equation}
where the slope has a canonical value of $q\sim4.8$ derived by Fraser \& Kavelaars (2009). The authors of this work also predicted a shallower slope for objects with diameters smaller than $D_b=60$ km, and this $D_b$ is set to be the minimum diameter $D_{min}$ of our synthetic samples. This considerably alleviates the computational time for the Monte Carlo simulations. By adopting $M_{TSCs}=0.02M_{\oplus}$, the synthetic samples have a total mass of $\sim0.0125M_{\oplus}$, with an extension from $D_{max}=800$ km to $D_{min}=60$ km. Under this condition, we generate $N_{MC}=122,000$ synthetic samples with assigned diameters from Eq. (\ref{sizeDistribution}). Accordingly, their masses are calculated by Eq. ({\ref{mass}}), and a random mass uncertainty within 10\% is introduced for each one. We note that, for a given $M_{TSCs}$, the size distribution of the TSCs is independent of the location (Gladman et al. 2001). Then even the power-law slope $p$ has a break around $D_b$, the spatial mass distribution of synthetic samples could remain nearly the same for a smaller $D_{min}$ (e.g., down to 1 km). Consequently, the Monte Carlo measurements of the total angular momentum would not be strongly affected.

We now turn to assign orbital elements to the above synthetic samples. It is obvious that the deviation of the total angular momentum direction of the minor planets from that of the Sun-planets system should be due to the inclined components. So, we first assume an unbiased inclination distribution of the TSCs with the form (Brown 2004)
\begin{equation}
f(i)\propto\sin i [C \exp(-i^2/2\sigma_1^2)+(1-C)\exp(-i^2/2\sigma_2^2)],
\label{iDistribution}
\end{equation}
where the parameters $C=0.83$, $\sigma_1=2.6$, $\sigma_2=16$, and the inclination $i$ is relative to the invariable plane of the Sun-planets system. Since Fraser \& Kavelaars (2009) found no difference in the size distribution (Eq. (\ref{sizeDistribution})) between the low- and high-inclination objects, for each synthetic sample with assigned mass, we chose the inclination randomly using the function $f(i)$. Then, as in Volk \& Malhotra (2017), the semimajor axis $a$ of a synthetic sample is chosen randomly in the range
\begin{equation*}
a_{real}-\Delta a < a < a_{real}+\Delta a,
\end{equation*}
where $a_{real}$ is the semimajor axis of a randomly selected real TSC, the values of $\Delta a$ are $0.01a_{real}$ and $0.05a_{real}$ for $a_{real}<50$ AU and $a_{real}\ge50$ AU, respectively; and the eccentricity $e$  is chosen randomly in the range
\begin{equation*}
e_{real}-\Delta e < e < e_{real}+\Delta e,
\end{equation*}
where $e_{real}$ is the eccentricity of another randomly selected real TSC, and the uniform value of  $\Delta e$ is $0.05e_{real}$. The other three orbital elements are all chosen randomly between $0^{\circ}$ and $360^{\circ}$. Thus in our Monte Carlo simulations, besides the observational incompleteness, we also implicitly take into account the mass and orbit uncertainties of observed objects.

In this way, we can measure the total angular momentum of the `complete' TSCs consisting of ten known TSCs with $D>800$ and a synthetic population of  $N_{MC}=122,000$ for all the smaller ones. This procedure has been repeated to perform 10,000 separate Monte Carlo simulations. We find that the average $\alpha_2$ of the 10,000 measurements is $0.8644^{\circ}$, and the $1\sigma$, $2\sigma$, and $3\sigma$ uncertainties that contain 68.2\%, 95.4\%, and 99.7\% of the simulated $\alpha_2$ have deviations of $0.2009^{\circ}$, $0.3991^{\circ}$ , and $0.5948^{\circ}$, respectively. This result shows that a small $\alpha_2$ for the TSCs can be reliable at the greater than $3\sigma$ level. Here, we intend to illustrate that the contribution from the unseen TSCs could be sufficient to induce a smaller $\alpha_2$, while only updated observations can tell us a precise value. However, we cannot fail to notice that the measured angles $\alpha_1$ from the 10,000 Monte Carlo simulations have an average value of $>8^{\circ}$, and the $1\sigma$, $2\sigma,$ and $3\sigma$ errors are $0.1563^{\circ}$, $0.3137^{\circ}$, and $0.4683^{\circ}$. This indicates that the contribution of the three dwarf planets, Eris, Pluto, and Haumea, to the total angular momentum of the TSCs is still significant even considering the observational incompleteness here.

We have to remark that the complete TSC population could possibly be more massive than $0.02M_{\oplus}$, for example, $\sim0.03M_{\oplus}$ (Booth et al. 2009) or a bit higher. This would yield more unseen TSCs and, correspondingly, would lower the proportion of the total angular momentum possessed by the dwarf planets. So, we can anticipate that the resultant $\alpha_1$ may continue to decrease. Similar Monte Carlo simulations for $M_{TSCs}=0.03M_{\oplus}$ have been performed by adopting a larger amount of $N_{MC}=220,600$ synthetic samples. We find that the simulated angles $\alpha_1$ for the TSCs achieve a smaller average value of $5.5427^{\circ}$ with a $1\sigma$ error of $0.1296^{\circ}$. The cases of even larger $M_{TSCs}$ are considered for the determination of the invariable plane of the solar system in Section 4.

We end by stressing that the key improvement in this paper is to compute the real angular momentum by introducing the parameter mass. Furthermore our results suggest that several known distant and most massive dwarf planets are likely to carry sufficient angular momentum to heavily influence the total angular momentum of the minor planets.

\subsection{Contribution of the most massive minor planets}

\begin{figure}
  \centering
  \begin{minipage}[c]{0.5\textwidth}
  \centering
  \hspace{0cm}
  \includegraphics[width=9cm]{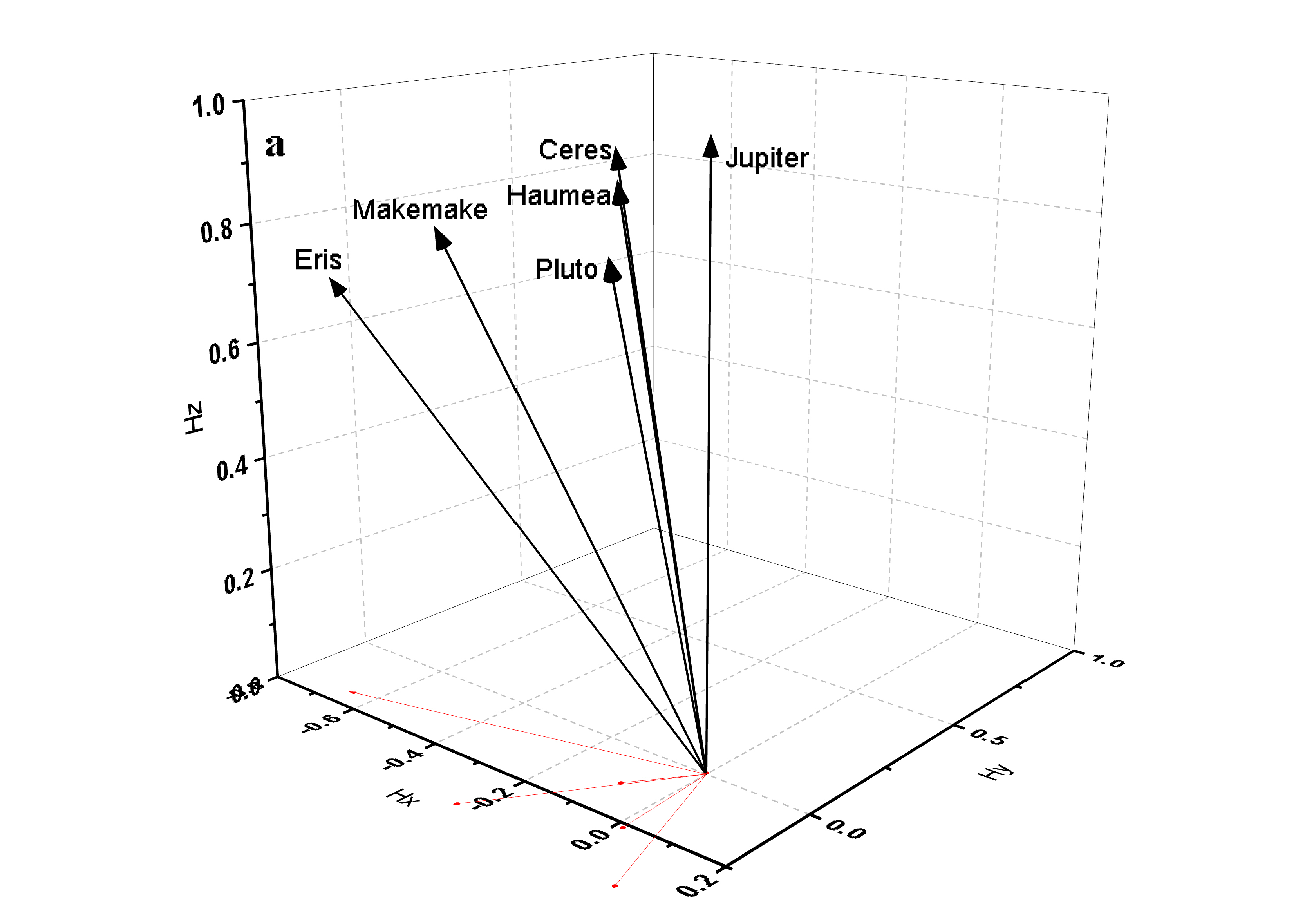}
  \end{minipage}
  \begin{minipage}[c]{0.5\textwidth}
  \centering
  \hspace{0cm}
  \includegraphics[width=9cm]{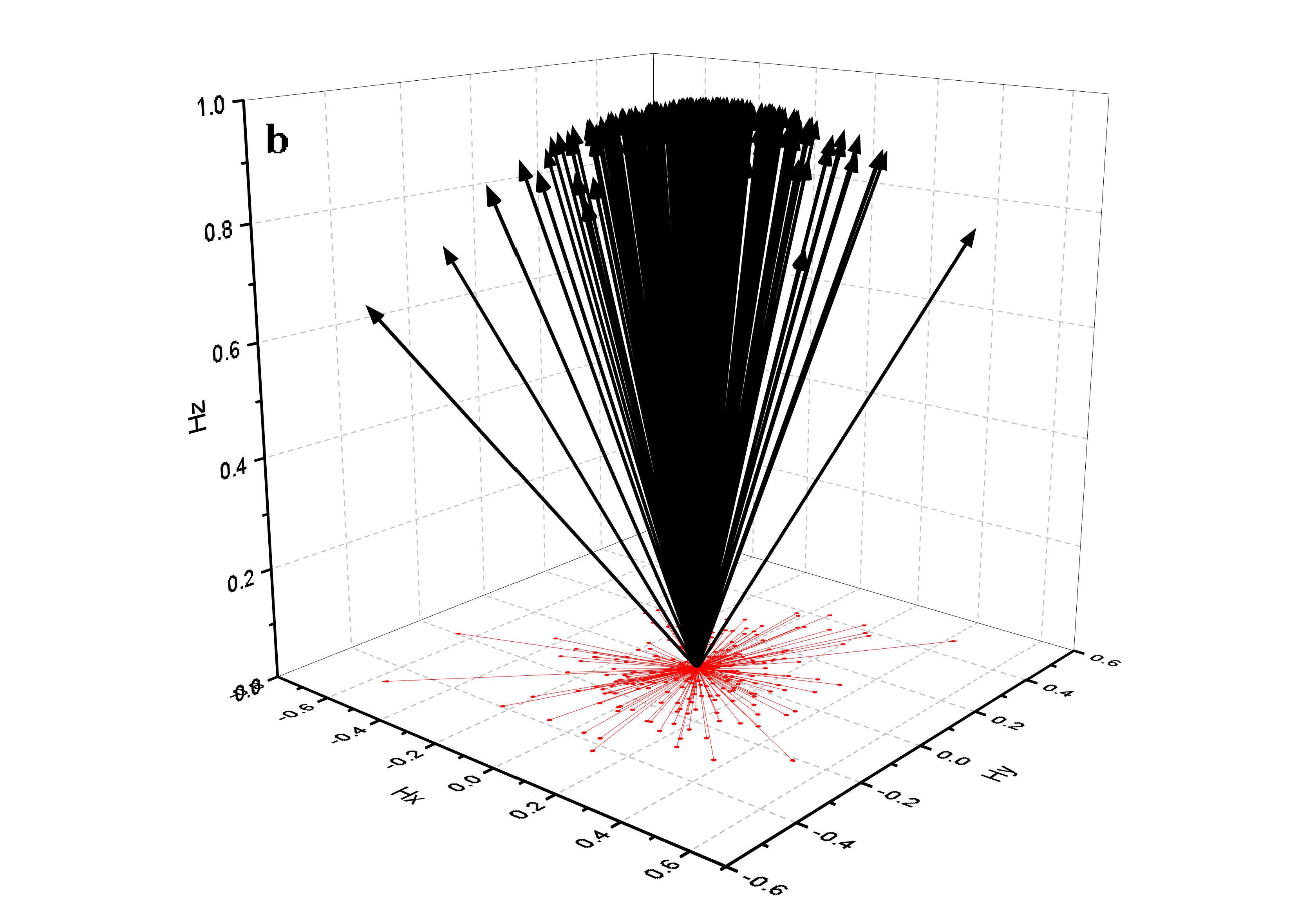}
  \end{minipage}
 \caption{Distribution of the angular momentum vectors of the most massive minor planets.  The reference plane $H_x$--$H_y$ is chosen to be the invariant plane of the Sun-planets system. For each object, the angular momentum vector $\vec{H}$ has been scaled by its module $|\vec{H}|$, that is, displaying the unit vector $\vec{\hat{H}}=\vec{H}/|\vec{H}|$, such that the several orders of magnitude difference in the lengths of arrows can be avoided. The red lines are the projections of $\vec{\hat{H}}$ on the $H_x$--$H_y$ plane. (a) For the five dwarf planets, together with Jupiter. (b) For the 279 minor planets with measured masses from Sect. 2.1.1.}
 \label{AMdistribution}
\end{figure} 

The five dwarf planets currently recognized by the International Astronomical Union (IAU) are all on highly inclined orbits, as noted in Fig. \ref{alpha_vari}. According to the results in Sect. 3.1, the value of $\alpha$ decreases significantly when the influence of the several most massive minor planets is removed in our calculations; the direction of the total angular momentum of the minor planets becomes nearly aligned with that of the Sun-planets system. So, the contribution of these massive objects needs to be studied in more detail.

Figure \ref{AMdistribution}a shows the distribution of the unit angular momentum vectors, $\vec{\hat{H}}=\vec{H}/|\vec{H}|$, of dwarf planets together with that of Jupiter. The use of unit vectors can eliminate the several orders of magnitude difference in the lengths of arrows, and give a clear comparison of their directions. For the sake of better display, in this subsection we choose the reference plane $H_x$--$H_y$ to be the invariable plane of the Sun-planets system. Such a transformation will not affect the relative angles among the angular momentum vectors. The projections of $\vec{\hat{H}}$ onto the reference plane (denoted by red lines) show that all the dwarf planets have negative vector components $H_y$, indicating somehow they happened to achieve similar orbital orientations. Since the dwarf planets possess a considerable fraction of the total angular momentum $\vec{H}_{minor}$ of the minor planets, their similar orbital orientations can lead to a substantial deviation of $\vec{H}_{minor}$ from $\vec{H}_{S\&p}$, that is, the large relative angle $\alpha_1$ between these two vectors. In Fig. \ref{AMdistribution}a, the inclusion of Jupiter's unit angular momentum vector, which is close to the direction of $\vec{H}_{S\&p}$ (i.e., $H_z$-axis), is to visualize the magnitude of $\alpha_1$. 

For the dwarf planets, the similar $\vec{\hat{H}}$ orientations toward the negative direction of the $H_y$-axis are not a consequence of reference plane selection, but indeed reflect the asymmetrical distribution of their angular momenta. When expanding the sample to the 279 relatively large minor planets with measured masses (see Sect. 2.1.1), we find that the $\vec{\hat{H}}$ distribution on the $H_x$--$H_y$ plane looks homogeneous, as shown in Fig. \ref{AMdistribution}b. Therefore, after excluding the dwarf planets with peculiar angular momenta (large moduli, similar directions), the vector $\vec{H}_{minor}$ could be much closer to the the $H_z$-axis, coinciding with the direction of  $\vec{H}_{S\&p}$. From this point of view, the reason for the steep decrease of $\alpha$ by removing the three most massive minor planets, as shown in Fig. \ref{alpha_vari}, becomes quite apparent.

One should notice that, due to the mentioned secular perturbations by the planets, the angular momentum vectors of the dwarf planets can process on the timescale about 10 Myr, and the corresponding directions would be modified. Therefore, the relative angle $\alpha_1$ between the total angular momenta of the dwarf planets and the Sun-planet system could have a certain amount of change over time. To evaluate this change, we performed numerical orbital integrations of the dwarf planets under the influence of the Sun and planets. During the 10 Myr evolution, the angle $\alpha_1$  varies in the range of 10$^{\circ}$-30$^{\circ}$. This suggests that, as long as the contribution of the dwarf planets is taken into account in the total angular momentum calculation for the minor planets, the angle $\alpha_1$ could always keep a large value.


\section{Invariable plane of the solar system}

\begin{table*}
\centering
\begin{minipage}{17cm}
\caption{Inclination $i_m$ and the ascending node $\Omega_m$ of the invariable plane of the solar system in the barycentric frame with respect to the J2000.0 ecliptic plane, measured at the epoch of 2016 November 8. The top row refers to the basic system (\Rmnum1) including the Sun, the eight planets, and Pluto; below are the more complete systems (\Rmnum2--\Rmnum6) that have taken into account the effects of individual minor planet populations. The difference $\Delta i_m$ between each of the systems \Rmnum2--\Rmnum6 and the system \Rmnum1 is  presented, and with the displacement $\Delta r_b$ of the solar system's barycenter accordingly.}      
\label{ip}
\begin{tabular}{ l l c c c c}       
\hline                 
System                        & Minor planet population                                                            &                           $i_m$                                                    &             $\Omega_m$                                          &      $\Delta i_m$(mas)   &      $\Delta r_b$  (km)  \\

\hline
 
\Rmnum1 (basic)    &   Pluto                                                                                             &              $1^{\circ}34'4\mathbf{3''.33124}$           &       $107^{\circ}3\mathbf{4'56''.17710}$        &               --                            &          --   \\

\Rmnum2                  &   Pluto, Ceres, Pallas, Vesta (PCPV)                                         &              $1^{\circ}34'4\mathbf{3''.34059}$          &       $107^{\circ}\mathbf{34'56''.21287}$        &           9.35                           &          0.25  \\
 
\Rmnum3                  &   PCPV, Eris, Haumea, Makemake                                           &              $1^{\circ}34'4\mathbf{4''.59456}$           &       $107^{\circ}3\mathbf{3' 28''.97049}$        &         1263                          &          96.51   \\

\Rmnum4                  &  279 samples with measured masses                                      &              $1^{\circ}34'4\mathbf{4''.36847}$           &       $107^{\circ}3\mathbf{3'26''.43100}$         &         1037                          &           112.61 \\

\Rmnum5                  &  135325 samples with measured masses\&diameters       &              $1^{\circ}34'4\mathbf{4''.37855}$           &       $107^{\circ}3\mathbf{3'26''.92827}$         &         1047                         &             111.38 \\

\Rmnum6                  &   718041 nominal samples                                                           &              $1^{\circ}34'4\mathbf{4''.37736}$           &       $107^{\circ}3\mathbf{3'26''.97123}$       &         1046                         &             111.27 \\            

\hline
\end{tabular}
\end{minipage}
\end{table*}

A natural application of our angular momentum study is to determine the invariable plane of the solar system. Souami \& Souchay (2012) showed that the inclusion of Pluto and Ceres is essential to calculate the orientation of the invariable plane. Considering the more important contributions of the other three dwarf planets to the total angular momentum of the minor planets, a further examination of this plane is necessary.

In order to evaluate and update the orientation of the solar system's invariable plane, we take into account the additional effects of: (1) the three dwarf planets Eris, Haumea, and Makemake; (2) the numerous nominal samples with assigned masses from our database. Here, the invariable plane is defined by the total angular momentum of the Sun, the eight planets, and a certain minor planet population.

The inclination $i_m$ and the ascending node $\Omega_m$ of the invariable plane is computed in the barycentric frame referred to the J2000.0 ecliptic plane, at the epoch of 2016 November 8. For comparison, we firstly re-did the measurements of $i_m$ and $\Omega_m$ for the two systems considered in Souami \& Souchay (2012): one is a basic system (\Rmnum1) that consists of the Sun, the eight planets, and Pluto; the other is a more complete system (\Rmnum2), to which the three MBAs Ceres, Pallas, and Vesta are added. The results are shown in the top two rows of Table \ref{ip}. We observe that the inclination difference $\Delta i_m$ between systems \Rmnum1 and \Rmnum2 is 9.35 mas, which tallies well with the value of 9.14 mas reported in Souami \& Souchay (2012). Over the 100 yr interval in our numerical integrations, the variation of $i_m$ is on the order of $10^{-9}$ mas. Then the $\sim0.21$ mas difference in $\Delta i_m$ probably arises from the initial conditions of Ceres, Pallas, and Vesta in the ASTORB file. Nevertheless, comparing to the absolute value of $\Delta i_m$, this magnitude of the difference is sufficiently small, and the validity of our calculation procedure has been verified.

Then we are able to further investigate the influence of the other minor planets on the invariable plane of the solar system, in addition to Pluto, Ceres, Pallas, and Vesta (PCPV for short). After introducing the three dwarf planets Eris, Haumea, and Makemake, we find that system \Rmnum3 has a significant increase in $\Delta i_m$ up to 1263 mas, as shown in Table \ref{ip}. Such an $i_m$ difference is two orders of magnitude larger than that for system \Rmnum2 including only PCPV in Souami \& Souchay (2012). Thus we propose that these three big bodies must be taken into account to determine the invariable plane.  Indeed, the contributions from minor planets to the variation of the invariable plane are linearly addible, that is, we have 
\begin{equation}
\Delta i_m=\Delta^{(1)} i_m+\Delta^{(2)} i_m+\cdots+\Delta^{(n)} i_m+\cdots,
\label{im}
\end{equation}
where the number in parentheses refers to the $n$-th sample. Then more specifically, the $i_m$ difference caused by the inclusion of Eris, Haumea, and Makemake is $\Delta^{(EHM)} i_m\sim1254$ mas. It is important to point out that the quantified effect of certain minor planet(s) on the invariable plane, for example, $\Delta^{(EHM)} i_m$, is independent from any other additional contributions to the solar system's angular momentum.

Finally, we evaluate the inclinations $i_m$ of the invariable plane for the even more complete systems \Rmnum4, \Rmnum5, and \Rmnum6, which include the 279 minor planet samples with measured masses, the 135325 samples with measured masses or diameters, and all the  718041 nominal samples, respectively. Although these systems may have uncertainties in the angular momenta of the faint minor planets, the results show that the $i_m$ differences with respect to the basic system \Rmnum1 are always greater than 1000 mas, that is, over 100 times larger than that for system \Rmnum2. According to Eq. (\ref{im}), besides system \Rmnum3 samples, the additional $i_m$ contribution from the rest of the minor planets in the solar system could be around $\Delta^{(rest)} i_m=-220$ mas. The negative value of $\Delta^{(rest)} i_m$ indicates that the deviation of the invariable plane due to Eris, Haumea, and Makemake could be somewhat compensated by numerous smaller objects, but not completely.

One must bear in mind that the observational incompleteness of the TSCs can considerably affect the total angular momentum of the minor planets. To explore the magnitude of the variation of the invariable plane for the intrinsic minor planet population, we calculate the values of $\Delta^{(rest)} i_m$ using the Monte Carlo simulations constructed in Sect. 3.2. Among nominal minor planets, we replace the known TSCs with the `complete' TSCs, which are composed of ten real samples with $D > 800$ km and hundreds of thousands of synthetic samples with $D\le800$ km. We allow the total mass $M_{TSCs}$ of the `complete' TSCs to vary from $0.02M_{\oplus}$ to a likely upper limit of $0.1M_{\oplus}$ (Luu \& Jewitt 2002; Vitense et al. 2010). It is interesting to find that, for any $M_{TSCs}$ within the above mass range, the 10,000 Monte Carlo simulations always give $\Delta^{(rest)}\sim-246$ mas with $1\sigma$ errors $\sim50-140$ mas. We therefore conclude that the orientations of the measured invariable plane using systems \Rmnum4, \Rmnum5, and \Rmnum6 (listed in Table \ref{ip}) are within $1\sigma$ of that of the intrinsic invariable plane. Furthermore, the deviation of our invariable plane from the recent measurement by Souami \& Souchay (2012) is on the order of $\Delta i_m=1000$ mas, which is quite robust and not due to observational selection.

Besides the orientation of the invariable plane, the solar system's barycenter may also slightly move after introducing different minor planet populations. Generally, the barycenter is derived from system \Rmnum1 of only the Sun, the planets, and Pluto, as in the Jet Propulsion Laboratory (JPL) planetary ephemerides. Here, we further assess its relative displacement in the position, denoted by $\Delta r_b$, for systems \Rmnum2-\Rmnum6. As shown in the last column of Table \ref{ip}, the values of $\Delta r_b$ are smaller than or around 100 km. We have confirmed that all the obtained results are fairly insensitive to this tiny location uncertainty of the solar system's barycenter.


\section{Conclusions and discussion}

The contribution of the minor planets to the total angular momentum of the solar system has not been fully studied before, especially taking into consideration that a large number of them are observed on high-inclination orbits. In this paper, we have constructed a database containing orbital elements and reasonable masses of 718041 minor planets, including NEOs, MBAs, JTs, and TSCs.  We then carried out, for the first time, an accurate measurement of the total angular momentum of the known minor planets, $\vec{H}_{minor}$, acquiring a modulus of $|\vec{H}_{minor}|=1.7817\times10^{46}$ g $\cdot$ cm$^2$ $\cdot$ s$^{-1}$.  By comparing $\vec{H}_{minor}$ to the total angular momentum $\vec{H}_{S\&p}$ of the Sun-planets system, we obtained the relative angle $\alpha$ between these two vectors. This approach yields a value of $\alpha=\alpha_1=14.74^{\circ}$. The result seems contradictory to the expectation that, even if there was asymmetrical change in the overall angular momentum of these small bodies, this change would be distributed to the entire solar system due to secular perturbations by the planets, and the angle $\alpha$ would decrease to a rather small value in $\sim10$ Myr (Volk \& Malhotra 2017). Our numerical experiments indicate that the upper limit of $\alpha$ should be around $4^{\circ}$.

We then demonstrated that such a large deviation between the directions of $\vec{H}_{S\&p}$ and $\vec{H}_{minor}$ is due to the contamination of the massive dwarf planets. They are not only moving on high-inclination orbits, but also have similar vectorial directions of the angular momenta. Thus the direction of the vector $\vec{H}_{minor}$ can be highly tilted. As long as we exclude the largest three dwarf planets, Eris, Pluto, and Haumea, the corresponding angle $\alpha=\alpha_2$ between the vectors $\vec{H}_{S\&p}$ and $\vec{H}_{minor}$ drops sharply to only about $1.76^{\circ}$. It suggests that the remaining minor planets have a direction of the total angular momentum very close to that of the Sun-planets system, despite any possible external perturbation in the past. In fact, the selection of any individual minor planet group (e.g., TSCs) will give a similar result, represented by the substantially different values of $\alpha_1$ and $\alpha_2$ in Table \ref{alpha12}. The mass and orbit uncertainties of the minor planets, and the observational incompleteness, have also been considered for the calculations of the angle $\alpha_2$. The results confirm that the value of $\alpha_2$ is intrinsically small, and the contribution of the dwarf planets to the total angular momentum of the minor planets is significant.

As an application of our angular momentum study, we evaluated the orientation of the invariable plane of the solar system by taking into account the minor planets. We showed that the influence of the three dwarf planets Eris, Haumea, and Makemake is very significant, inducing a difference of 1254 mas in the inclination $i_m$ of the invariable plane, while in Souami \& Souchay (2012) they found an $i_m$ difference of only 9 mas due to the inclusion of Ceres, Vesta, and Pallas. Moreover, the effects of the other minor planets (including the unseen ones) have also been estimated; collectively these small bodies add a $1\sigma$ uncertainty of $50-140$ mas to $i_m$. Thus we conclude that the consideration of Eris, Haumea, and Makemake is vitally important to determine the orientation of the solar system's invariable plane.

A by-product of the contribution of the angular momenta of the minor planets relates to the issue of the current tilt $\delta\sim6^{\circ}$ between the invariant plane of the Sun-planets system and the solar equator (Beck \& Giles 2005). According to planetary system formation theories, planets originated from the solar nebula that was coplanar with solar equator, thus their total angular momentum vector was approximately parallel to the rotation axis of the Sun, meaning the tilt $\delta$ should be close to zero. More accurately, the definition of $\delta$ ought to be revised to consider the invariant plane of the whole planetary system, by introducing the minor planets that were also born in the solar nebular. Unfortunately, even if all the 718041 samples in our database were added to the planetary system, the invariant plane would not change by more than $0.001^{\circ}$. This slight shift certainly cannot account for the $\sim6^{\circ}$ tilt, and other mechanisms have to be involved, such as the unseen Planet 9 with about ten Earth masses and semi-major axis of hundreds of AU (Bailey et al. 2016; Lai 2016; Gomes et al. 2017).

For the total angular momentum of the solar system, the contribution from the planetary moons has not yet been taken into account. An estimation could be done by adding the masses of the moons to each individual host planet. We consider the largest moons with masses on the order of $10^{23}$ kg in the outer solar system, which are Io, Europa, Ganymede, Callisto (moons of Jupiter), Titan (moon of Saturn), and Triton (moon of Neptune). We find that, comparing with the values of $\alpha_2$ shown in Table \ref{alpha12}, the resultant variations are always smaller than $0.0001^{\circ}$ for different minor planet populations. This is easy to understand because a moon can hardly affect the angular momentum of its host planet. Nevertheless, the total angular momentum of these six moons is calculated to be $6.3579\times10^{46}$ g $\cdot$ cm$^2$ $\cdot$ s$^{-1}$, which is about 3.6 times larger than that of the known minor planets (i.e., $|\vec{H}_{minor}|$).

Until now, we have only been concerned about the orbital angular momentum for the celestial bodies in the solar system, while the spin angular momentum has been neglected. For the minor planets, since nearly none of the known samples with absolute magnitudes $H<22$ have rotation periods greater than 2.2 hours (Pravec et al. 2002), we estimate that the upper limit of their total spin angular momentum is less than $10^{-8}$ of the total orbital angular momentum. Therefore, the neglect of the spin angular momenta of the minor planets is quite straightforward. As for a planet, the orbital angular momentum is at least five orders of magnitude larger than the spin angular momentum. Simple calculation shows that, including the spin angular momentum or not, the direction of the total angular momentum of the planets would not shift by more than $0.0001^{\circ}$, thus the angle $\alpha_2$ would be about the same. We note that the total spin angular momentum of the planets is $\lesssim0.8\times10^{46}$ g $\cdot$ cm$^2$ $\cdot$ s$^{-1}$, smaller than but still comparable to the obtained $|\vec{H}_{minor}|$. Combining the contribution from the planetary moons, in the future, we intend to use these two sources of angular momentum to determine the orientation of the invariable plane of the solar system.

\begin{acknowledgements}
This work was supported by the National Natural Science Foundation of China (Nos. 11973027, 11473015, 11933001, and 11473016), and the Fundamental Research Funds for the Central Universities (No. 020114380024). JL has to thank the China Scholarship Council for the funds to pursue study at Northwestern University as a visiting scholar in the year 2016-2017. The essential part of this research was done at that time. The authors are grateful to Dr. Xiaosheng Wan for the discussion. We would also like to express our thanks to the referee for the valuable comments, which helped to considerably improve this paper.

\end{acknowledgements}

\end{document}